\documentclass[11pt,aps,prl,twocolumn,reprint]{revtex4-1}
\usepackage{graphicx}
\usepackage{amsmath}

\begin{document}
\title{Quantum single electron solitons near metal surface }

\begin{abstract}
A possibility of a quantum single electron soliton (QSES) formation in structures with different dimensionality (0, 1, 2, and 3D) and spectrum (parabolic and linear) placed near metal surface is discussed. They originate as solutions of the nonlinear Schr{\"o}dinger equation allowing for interaction with image charges in metal. The binding energy of those quasi-particles could exceed the thermal energy at room temperature. 
\end{abstract}
\author{V. Vyurkov, D. Svintsov}
\affiliation{Institute of Physics and Technology RAS}
\affiliation{Moscow Institute of Physics and Technology}

\maketitle

Here we consider a possibility of a quantum single electron soliton (QSES) formation in various structures placed near a metal electrode (gate). It originates as a solution of the nonlinear Schr{\"o}dinger equation, therefore, the term QSES arises. At the same time, with regard to their physical nature (due to polarization of environment) they could be also called as polarons. Previously, a self-consistent solution of the Schr{\"o}dinger and Poisson equations (mean-field or Hartree approximation) was widely used for simulation of quantum many-particle systems, for example, field-effect transistors (see~\cite{Quantum_SOI} and references herein). However, in the case the Coulomb repulsion between particles indispensably dominates, hence, a soliton formation is prohibited. For a single electron that repulsion, evidently, vanishes. Therefore, in the situation the straightforward self-consistent procedure has to be substantially modified to exclude the repulsion of an electron by its own field. Meanwhile, in present communication we discuss only simple structures with an explicit description. 

First of all, in the case of a single electron, the obstacle appears how to introduce an image-charge potential into the Schr{\"o}dinger equation. The potential created by a trial point charge undoubtedly fails, at least, for two reasons. On one hand, that potential does not exist without charges; on the other hand, the question - what is a charge distribution on the metal surface corresponding to that potential? - has no answer. The first proposal to cope with this problem was put forward in Refs.~\cite{Image-force-1,Image-force-2}. The authors suggested that the surface charge in metal distributes itself in response to the 'instantaneous' charge density of an electron $e|\Psi ({\bf r},t)|^2$. Lately, this hypothesis was to some extent justified in~\cite{Woodford,Filippov}. This approach was already used for description of an atom near a metal surface~\cite{Hydrogen-metal-plate} and damped Rabi oscillations of a charge qubit due to Joule losses in nearby metal electrodes (gates)~\cite{Filippov-2}. Here we adopt the approach used in those papers.
\begin{figure}
\center{\includegraphics[width=0.85\linewidth]{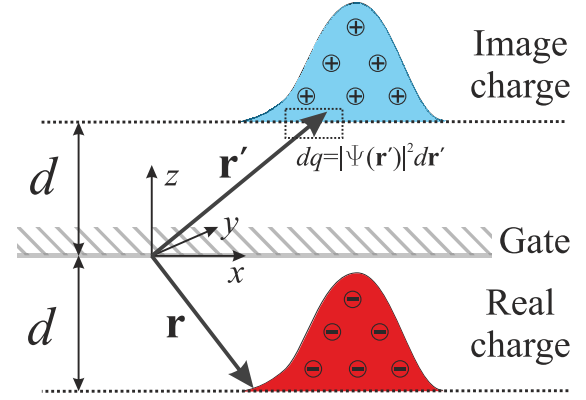}}
\caption{\label{Auxilary} Schematic view of electron probability density near metal gate and density of image charge}
\end{figure}

Then the potential energy originating from interaction of the electron with image charges in a metal plate (see Fig.~\ref{Auxilary}) is
\begin{equation}
V_C({\bf r},t)=-\frac{e^2}{2\kappa}\int{\frac{ \left| \psi ({\bf r'},t)\right| ^2 d{\bf r'}}{\left|{\bf r} - {\bf r'}\right|}},
\end{equation}
the integration is performed over all image charges, $\left|{\bf r} - {\bf r'}\right| = \left[ (x-x')^2+(y-y')^2+4d^2\right]^{1/2}$ is the distance between image charge and point ${\bf r}$ in the channel, $d$ is the distance to the gate, and $\kappa$ is a dielectric permittivity. This term should be substituted into the Schr{\"o}dinger equation which becomes thus nonlinear:
\begin{equation}
\label{NLSE}
i\hbar\frac{\partial \Psi ({\bf r},t)}{\partial t} = -\frac{\hbar^2}{2m^*}\Delta\Psi({\bf r},t)-\frac{e^2 \Psi({\bf r},t)} {2\kappa } \int{\frac{ \left| \Psi ({\bf r'},t)\right| ^2 d{\bf r'}}{\left|{\bf r} - {\bf r'}\right|}},
\end{equation}
here $m^*$ is effective mass of electron. Of course, this equation must be accompanied with a normalization of the wave function per one electron.

For the soliton creation the particular form of $V_C({\bf r},t)$ is not important. The crucial property is the negative sign and that this term increases rapidly as the soliton shrinks. Figuratively speaking, a soliton is born in counteraction of potential and kinetic energies.

While an analytical solution of~(\ref{NLSE}) seems hardly possible, one can estimate the size and energy of solitons using a direct variational principle. Assuming some localized wave function $\Psi({\bf r},a)$ ($a$ is soliton size) as approximate solution of~(\ref{NLSE}), one minimizes the total energy of state with respect to $a$:
\begin{equation}
\label{Functional}
E(a)=\frac{\hbar^2}{2m}\int{|\nabla \Psi({\bf r},a)|^2 d{\bf r}}-\frac{e^2}{2\kappa}\int{\frac{|\Psi({\bf r},a)|^2|\Psi({\bf r'},a)|^2 d{\bf r } d{\bf r'}}{|{\bf r}-{\bf r'}|}}.
\end{equation}
Previously, such approach was successfully used to calculate the self-energy and the size of polarons in ionic crystals~\cite{Pekar}, where the phenomenon of self-localization also occurred..

\begin{figure}
\center{\includegraphics[width=0.85\linewidth]{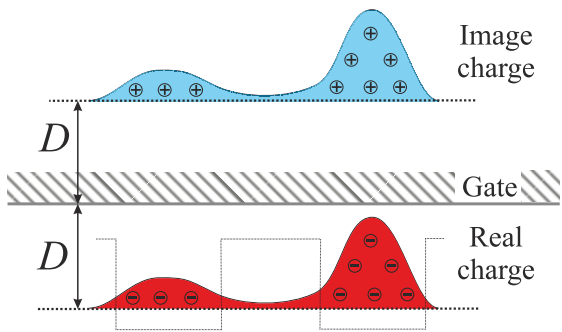}}
\caption{\label{DQD-1} Image charges in a metal plate acting on a single electron motion in a double quantum dot}
\end{figure}

\begin{figure}
\center{\includegraphics[width=0.85\linewidth]{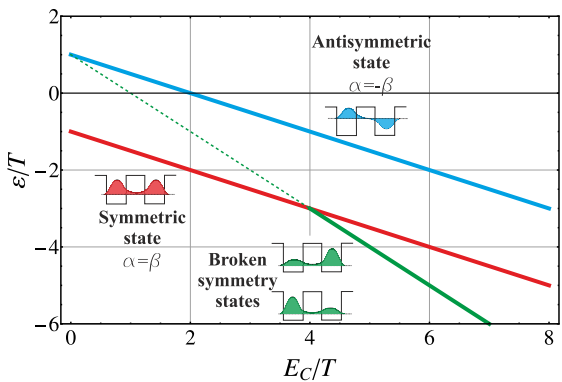}}
\caption{\label{DQD-2} Energies of stationary states in DQD vs. the ratio $E_C/T$ and corresponding distributions of probability density}
\end{figure}

We first consider the simplest case of quantum system described by Eq.~(\ref{NLSE}), namely, two identical quantum dots (a double quantum dot, DQD) with a single electron. The dots are placed near metal surface (Fig.~\ref{DQD-1}). This structure is promising for implementation of charge-based qubits. The nonlinear rate equations for probability amplitudes of finding electron in the left and in the right quantum dots ($\alpha$ and $\beta$, respectively) read
\begin{gather}
\label{Rate-Eqs}
i\hbar \frac{\partial \alpha}{\partial t} = -E_C |\alpha|^2 \alpha - T \beta,\\
i\hbar \frac{\partial \beta}{\partial t} = -E_C |\beta|^2 \beta - T \alpha,
\end{gather}
where $T$ is the tunneling coupling between dots, and $E_C$ is the energy of Coulomb interaction with metal.  Here $T$ and $E_C$ are analogous to kinetic and potential energy, respectively. The normalization $|\alpha|^2+|\beta|^2$ is observed automatically. 

Let us examine the stationary solutions of the system (\ref{Rate-Eqs}) assuming $\alpha \propto \beta \propto e^{i E t/ \hbar}$. For small parameters $E_C/T$ there are two energy levels
\begin{equation}
\varepsilon_{\pm} = \pm T - E_C/2,
\end{equation}
providing the lower one corresponds to symmetric, and the higher one to antisymmetric state. The situation drastically changes when the parameter $E_C/T$ exceeds its critical value equal to 4. A twice degenerate level with the energy 
\begin{equation}
\varepsilon = T - E_C
\end{equation}
appears, which corresponds hereafter to the ground state of the system (see Fig.~\ref{DQD-2}). In one of degenerate states the electron is primarily localized in the left dot, while in the other it is localized in the right dot. For $E_C > 4T$, the amplitude of Rabi oscillations in the qubit becomes very small~\cite{Lipki}.

This result holds valid for any number of identical quantum dots. The localization of electron in a certain dot for strong Coulomb interaction could be treated as formation of a soliton. Degenerate solutions localized near identical minima of potential energy were obtained mathematically in \cite{Nonspreading-wave-packets} (without any regard to the origin of nonlinearity in the Schr{\"o}dinger equation, though). 

It is not clear yet whether those localized states can be used for quantum computation. What is now clear for sure is that the interaction with surrounding media via its polarization even when it is rather small ($E_C/T < 1$) has a strong influence on a phase evolution of qubit. Worth noting this action cannot be anyhow compensated because during computation the quantum dot populations remain unknown. It could be a general drawback of almost all charge qubits. Fortunately, the quantum computer based on space states without charge transfer was recently proposed~\cite{Gorelik}. It allows avoiding all harmful processes related to moving charges.

For continuous wave function $\Psi(x,t)$ the nonlinear term becomes more complicated as the potential energy depends on the distribution of image charges. However, for electrons localized in a quantum wire it can be roughly approximated as follows 
\begin{equation}
V_C(x,t) \approx -\frac{e^2}{\kappa }|\psi (x,t){{|}^{2}}\ln (a/d),
\end{equation}
where $a$ is soliton size. On omitting this logarithmic factor which describes a weak correction depending on a soliton size, one arrives at the conventional representation of the 1D nonlinear Schr{\"o}dinger equation
\begin{equation}
\label{Schrodinger-1D}
-\frac{\hbar^2}{2m^*}\frac{d ^2\Psi }{dx^2}-\frac{e^2}{\kappa}|\Psi|^2 \Psi =i\hbar \frac{\partial \Psi }{\partial t}.
\end{equation}

It should be noted that according to a solution of Poisson equation for many-electron system the second term (potential energy) in the left-hand side of Eq. (3) has a positive sign and is equal to 
$+e^2|\Psi|^2\Psi/C$, where $C=\frac{\kappa}{\ln (d/2 r_0)}$ is specific capacitance of the wire with respect to metal, $r_0$ is the wire radius. The positive sign just means that a Coulomb repulsion between electrons dominates and hampers soliton formation.

The general solutions of the equation (\ref{Schrodinger-1D}) could be derived via inverse scattering transforms~\cite{Zakharov-Shabat}; they represent both knoidal waves and localized (solitary) waves. The latter are 
\begin{equation}
\Psi \left( x,t \right)=\frac{\sqrt{2\alpha /\nu }}{\cosh \left[ \sqrt{\alpha }\left( x-\hbar kt/m \right) \right]}{{e}^{i\left( kx-Et/\hbar  \right)}},
\end{equation}
where $\alpha ={{k}^{2}}-2mE/{{\hbar }^{2}}$, $\nu =2m{{e}^{2}}/{{\hbar }^{2}}\kappa $.

The normalization of wave function per one electron leads to the condition $4\sqrt{\alpha }=\nu $, hence, the characteristic size of the soliton is $a={\alpha^{-1/2}}=4{a_B}$, where 
\begin{equation}
{a_B}=\frac{\hbar^2 \kappa }{m^* e^2}
\end{equation} 
is the Bohr radius. The energy spectrum of solitons is continuous:
\begin{equation}
E=\frac{{\hbar^2}}{2m}\left[ {k^2}-{{\left( \frac{1}{4 a_B} \right)}^{2}} \right]
\end{equation}

The form of potential energy in Eq.~(\ref{Schrodinger-1D}) is valid if only the soliton size exceeds the distance to the gate. Otherwise, it should be replaced by $-\frac{e^2}{4\kappa D}$, which is the energy of a point charge near metal surface. Consequently, a soliton size is restricted by the distance to the gate $d$ and, in fact, $a=\max \{4 a_B,d\}$. Worth noting in 1d a soliton always exists for parabolic spectrum of particles.

For parameters of a silicon nanowire ($m^*/m_e=0.2$) covered by SiO$_2$ ($\kappa =4$) as gate dielectric the binding energy of standing soliton is 10 meV, which is comparable with thermal energy at room temperature equal to 25 meV. The binding energy for heavy hole solitons in silicon ($m^*/m_e=0.5$) approaches this thermal energy. When metal is placed all-around a wire the binding energy could be augmented.

It should be emphasized that in spite of a resemblance of terms, the QSES just discussed above has no relation to a single-electron soliton (SES) in Ref.~\cite{Likharev-SES} where an electron probabilistically jumps along a one-dimensional chain of tunnel junctions.

For electrons localized in two dimensions, the nonlinear Schr{\"o}dinger equation in presence of metal reads: 
\begin{multline}
i\hbar \frac{\partial \Psi }{\partial t} = -\frac{\hbar^2}{2m^*}\left(\frac{\partial^2}{d x^2}+\frac{\partial^2} {dy^2} \right)\Psi \\ -\frac{e^2 \Psi}{2\kappa}\int\limits{\frac{|\Psi (x',y',t)|^2 dx'dy'}{\left[(x-x')^2+(y-y')^2+4{d^2}\right]^{1/2}}}.
\end{multline}

\begin{figure}
\center{\includegraphics[width=0.85\linewidth]{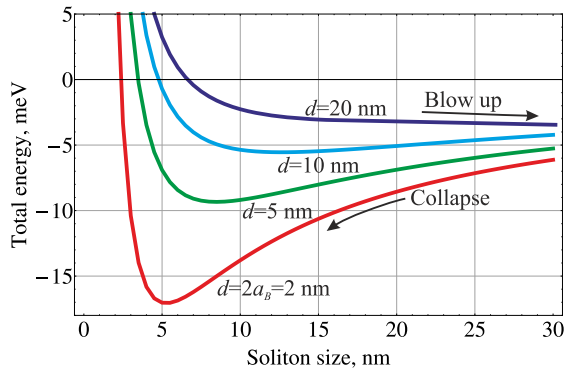}}
\caption{\label{2d_Energy} Energies of soliton in two-dimensional system as a function of its size $a$ for different distances to the gate $d$}
\end{figure}

The implementation of variational principle in this case leads to the results different from 1D case. Particularly, the soliton kinetic energy $K$ is inversely proportional to $a^2$:
\begin{equation}
K\propto \frac{\hbar^{2}}{m^* a^2},
\end{equation}
the similar dependence with an opposite sign holds for electrostatic energy $E_C$ provided $a\gg d$
\begin{equation}
E_C\propto -\frac{e^2 d}{a^2}.
\end{equation} 
The minimum energy depending on the ratio $K/E_C = a_B/d$ can be attained either at infinite soliton radius (when $a_B>d$), or at zero radius (when $a_b<d$). Such behavior of localized solutions is well-known in the problem of laser beam self-focusing~\cite{Townes-self-trapping}, where the equation for electric field amplitude analogous to the nonlinear Schr{\"o}dinger equation arises. Depending on the beam power, a soliton can either collapse due to the Kerr nonlinearity, or blow up due to diffraction. However, in our case of 2D electrons near metal surface the shrinkage to zero size is impossible virtue to a limited value of Coulomb interaction energy when $a/d\to 0$ (like for a point charge near metal surface): 
\begin{equation}
E_C\to \frac{e^2}{4\kappa d}.
\end{equation}
Therefore, the dependence of total soliton energy on its size can have a minimum at a finite value of $a$. The latter can be found from minimization of the functional 
\begin{gather}
E(a,d)=\frac{\hbar^2}{8 m d^2}\left\{ \left(\frac{2d}{a}\right)^2-\frac{32}{\pi}\frac{d}{a_B}\left(\frac{2d}{a} \right) I\left(\frac{2d}{a}\right)\right\},\\
I(x)=\int_0^\infty{dr_1 dr_2 \int_0^\pi{\frac{r_1 r_2 d\theta}{\sqrt{r_1^2 + r_2^2-2r_1 r_2 \cos\theta + x^2}}}};
\end{gather}
here we used the trial wave function $\Psi_{2d}(r)=(2/\pi a^2)^{-1/2}e^{-r/a}$ in Eq.~(\ref{Functional}). The dependence of soliton energy on its size $a$ and the distance to the metal is plotted in Fig.~\ref{2d_Energy}. It can be seen that a soliton formation is possible at $d \lesssim 10 a_B$, at larger distances to the metal a delocalized state is preferable with respect to minimum energy. For the layers of ultrathin silicon covered with SiO$_2$, one obtains $a_B = 1$ nm, $d \lesssim 10$ nm as a necessary condition for the soliton formation. 

In 3D a soliton formation seems still feasible in the vicinity of a semiconductor/dielectric interface close to the metal plate. In the case, the description requires a self-consistent solution of Schr{\"o}dinger and Poisson equations necessarily adapted to a single electron. 

Another localized quasi-particle (namely, a polaron) may also originate due to polarization of semiconductor (dielectric) environment. However, in the case, the description much differs from that for the metal plate. Indeed, although the energy of interaction of the electron with environment is still negative, but the energy of interaction between induced dipoles is positive and should be also taken into account. The interaction between dipoles can substantially diminish a polaron binding energy. However, this scrutiny is beyond the scope of present paper and could be found elsewhere.

It would be challenging to investigate the behavior of single-electron solitons in carbonic materials, like graphene and nanotubes. The dynamics of an electron in such system is described by the Dirac-like equation~\cite{Graphene-hamiltonian}, which can be easily generalized to account for interaction with metal surface 
\begin{multline}
\label{NLDE}
i\hbar \frac{\partial \vec{\Psi}({\bf r},t)}{\partial t} = v_F (\sigma_x \hat{p}_x + \sigma_y \hat{p}_y)\vec{\Psi}({\bf r},t) + \\
\frac{e^2 \vec{\Psi} ({\bf r},t)}{2 \kappa}\int{\frac{|\vec\Psi({\bf r'},t)|^2 d{\bf r'}}{|{\bf r}-{\bf r'}|}},
\end{multline}
here $\vec \Psi = \left\{\psi_A, \psi_B \right\}$ is a two-component wave function, $\sigma_i$ are Pauli matrices, $\hat p_i$ are momentum operators, and $v_F\simeq 10^6$ m/s is characteristic velocity of electrons (Fermi velocity). However, the solutions localized in both $x$- and $y$- directions for Eq.~(\ref{NLDE}) are impossible, which is closely related to the phenomena of chiral tunneling and Klein paradox~\cite{Chiral-tunneling}. Indeed, assuming a stationary solution $\vec\Psi\propto e^{-i\varepsilon t/\hbar}$ with a characteristic size $a$, Eq.~(\ref{NLDE}) in the limit of large distances $r\gg a$ gives rise to
\begin{gather}
i \frac{\partial \psi_A}{\partial r} = \frac{\psi_B}{r_0},\\
i \frac{\partial \psi_B}{\partial r} = \frac{\psi_A}{r_0},
\end{gather}
where $r_0 = \hbar v_F/\varepsilon$. The solutions of this system are harmonic functions of $r/r_0$. Thus, electron localization near metal gates in gapless materials looks impossible.

Nonetheless, in semiconductor nanotubes and graphene nanoribbons with parabolic spectrum and nonzero bandgap the soliton formation seems plausible. One more striking effect may then occur, namely, a spontaneous creation of electron-hole pairs   owing to spontaneous breaking of symmetry described above for two quantum dots. For that the bandgap should be sufficiently narrow. 

In conclusion, we discussed a possibility of a quantum single electron soliton (QSES) formation in different structures near a metal plate. Its binding energy can attain the thermal energy at room temperature and even exceed it. As those quasi-particles could be robust against scattering they seem promising for nanoelectronic applications.

The research was supported by the grants of Russian Academy of Sciences, Russian Foundation for Basic Research (11-07-00464) and Tohoku University (Japan).

\end{document}